\documentclass[
aps,%
12pt,%
final,%
notitlepage,%
oneside,%
onecolumn,%
nobibnotes,%
nofootinbib,%
superscriptaddress,%
noshowpacs,%
centertags]%
{revtex4}%
\usepackage{geometry}
\usepackage{graphics}

\begin{document}

\title{Color flows for the process \( gg\rightarrow B_{c}+c+\bar{b} \) }

\author{\firstname{A.~V.}~\surname{Berezhnoy}}
\email{aber@ttk.ru}
\affiliation{SINP MSU, Moscow, Russia}%

\begin{abstract}
The  contributions of different color flows 
into  the gluonic \(B_c\)-meson production cross section
has been calculated. This study is essential to simulate  
 \(B_c\)-meson production with the help of Pythia  program. 
The essence of matter is that in the frame work of the Lund model used by 
Pythia the hadronization way of the final partons 
and hadronic remnants depends on color flow type.  
The modified method for calculation
of the color flow contributions has been proposed. 
\end{abstract}

\maketitle

\section{INTRODUCTION}

A simulation of particle production processes at  modern   
hadronic experiments is practically impossible without Monte-Calro methods. 
The fact  of the matter is that the integro-differential 
equations, which are used to describe the evolution of initial partons and
the hadronization of  parton interaction products, are very complicated.
Recent time PYTHIA \cite{Pythia} is the  most dependable 
program for Monte-Carlo simulation of a particle production at high energies.
This software simulate all three studies, 
 on which the  hadronic production can be conventionally subdivided:
the initial parton evolution, the hard subprocess of  
the initial partonic interaction, 
the hadronization of the hard interaction products and 
hadronic remnants. The hard subprocess is calculated in the 
frame work of perturbative QCD. The hadronization in Phythia is described by
Lund model \cite{c_string}. In the frame work of this model
one supposes that outgoing color charges stretch the connecting force fields
(color strings), and it is these fields that eventually break up to produce the final state
hadrons. The force field is always stretched from a color triplet  to a color 
antitriplet. Color octets (for example, gluons) are 
treated as an excitation of the string.
A technique which allows to calculate the particular color flow contribution 
into the  cross section  has been performed in \cite{c_string}. 
Also  it has been demonstrated in \cite{c_string}
that at infinite color number limit \( N_{c}\rightarrow \infty  \) 
the interference terms between different color flows equal to zero. 

A wide set of standard hard processes can be simulated by  PYTHIA, however 
this set does not contain all processes, which would be interesting to study
at high energies.  That is why in the recent versions of PYTHIA  a user have a
possibility to include his own matrix element into the PYHTIA program.  
Therefore the problem of color flow separation appears. 
It is important to know the contribution of different color  flows into the 
cross section,  because each color flow corresponds to its sole 
hadronization way. 

In this work the color flows for the process of gluonic production of 
\( B_{c} \)-meson have been studied. The modified calculation method 
of the color flow contributions has been proposed, which is slightly differ
from  the traditional one.

It is worth to note that it is somewhat  difficult 
to calculate the cross section for the process of gluonic 
\( B_{c} \)-meson production, because \( B_{c} \)-meson production 
mechanism  can not be treated as \( b\bar{b} \)-pair production followed by
 \( b \)-quark hadronization into   \( B_{c} \)-meson. Therefore
\( B_{c} \)-meson production can not be described by 
the fragmentation function
\( b\rightarrow B_{c} \) (the fragmentation function calculations
have been  performed in \cite{fragmentation}).
It has been shown that under conditions of present-day and planed experiments 
the recombination mechanism of \( B_{c} \)-mesons production dominates.
To estimate the recombination contribution into the cross section 
one need to calculate  36 tree-level Feynman diagrams of 
order  \( O(\alpha ^{4}_{s}) \). These calculations have been done 
independently by several research groups 
\cite{Bc_Berezhnoy,Bc_Chang,Bc_Kolodziej,Bc_Baranov,
Bc_Slabospitsky,Bc_Sartogo}.
The results of  \cite{Bc_Berezhnoy,Bc_Chang,Bc_Kolodziej,Bc_Baranov} 
are in  a good agreement with each other.

An interest in \( B_{c} \)-meson production in hadronic interactions 
is increasing now due to the fact that LHC will have started to work in the 
nearest future. The last versions of PHYTIA based generator SIMUB \cite{SIMUB}
include the useful subroutines to simulate \( B_{c} \) production,
 which are based on codes of research groups \cite{Bc_Berezhnoy,Bc_Chang}. 
The fruitful scientific contacts between the author and SIMUB developers have 
caused the presented study.

\section{THE SEPARATION OF COLOR FLOWS}

\begin{figure}
\hspace*{-2.5cm}{\centering \resizebox*{1.2\textwidth}{!}
{\includegraphics{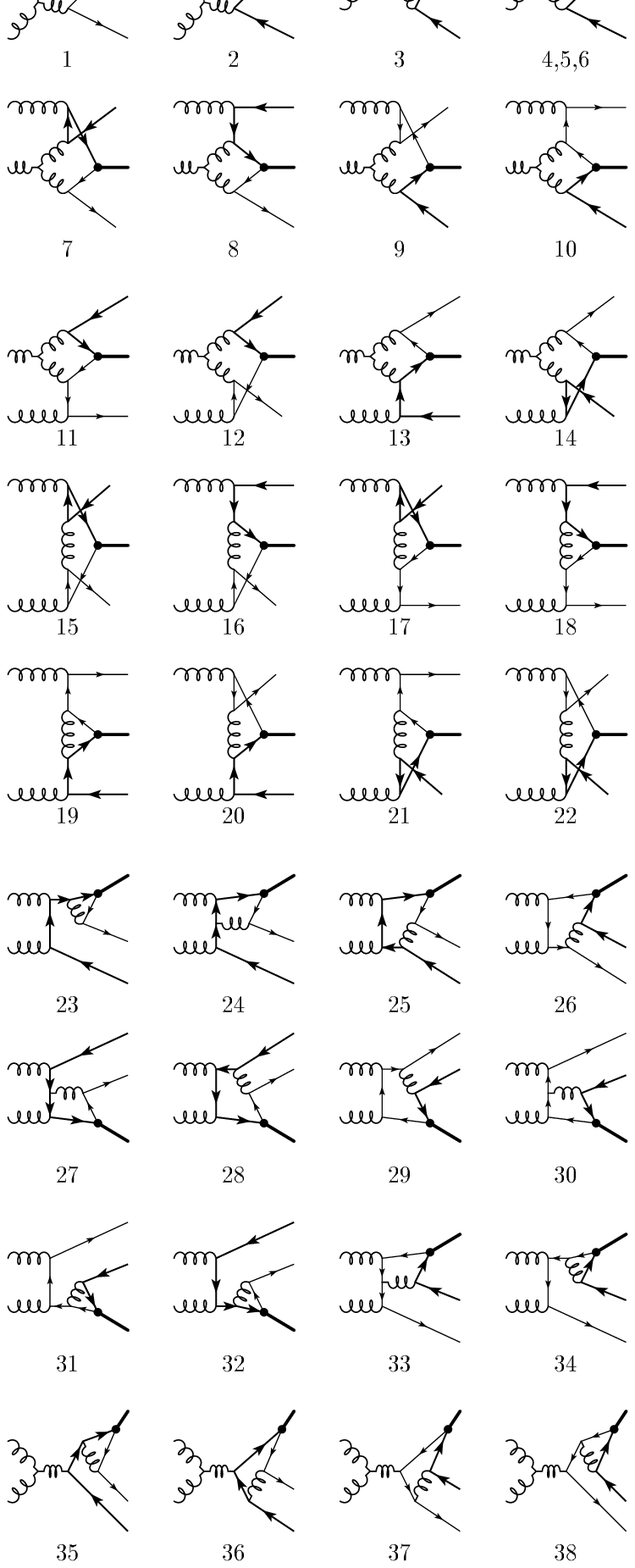}} \vspace*{-9cm}\par}

\setcaptionmargin{5mm}
\captionstyle{normal}
\caption{Feynman diagrams for 
\protect\( B_{c}\protect \)-meson production in 
the gluonic interaction.\hfill\label{diagr}}
\end{figure}

The Feynman diagrams for the process \( gg\rightarrow B_{c}+X \) are 
shown in Fig. \ref{diagr}. It is useful to treat 
the diagram with four gluonic vertex as three different diagrams, because that 
diagram contains three different color structure.
The color parts of the diagrams are given by following equations:
\[
T_{1}=f^{n_{1}g_{2}n_{2}}f^{n_{3}n_{2}g_{1}}t_{c\bar{c}}^{n_{1}}t_{b\bar{b}}^{n_{3}}\delta _{b\bar{c}},\]
 \[
T_{2}=f^{n_{1}g_{2}n_{2}}f^{n_{3}n_{2}g_{1}}t_{c\bar{c}}^{n_{3}}t_{b\bar{b}}^{n_{1}}\delta _{b\bar{c}},\]
\[
T_{3}=f^{n_{1}g_{1}g_{2}}f^{n_{2}n_{1}n_{3}}t_{b\bar{b}}^{n_{2}}t_{c\bar{c}}^{n_{3}}\delta _{b\bar{c}},\]
\[
T_{4}=f^{n_{1}g_{2}n_{2}}f^{n_{2}n_{3}g_{1}}t_{b\bar{b}}^{n_{3}}t_{c\bar{c}}^{n_{1}}\delta _{b\bar{c}},\]
\[
T_{5}=f^{g_{2}n_{1}n_{2}}f^{n_{2}n_{3}g_{1}}t_{b\bar{b}}^{n_{1}}t^{n_{3}}_{c\bar{c}}\delta _{b\bar{c}},\]
\[
T_{6}=f^{n_{1}n_{2}n_{3}}f^{n_{3}g_{2}g_{1}}t^{n_{1}}_{b\bar{b}}t^{n_{2}}_{c\bar{c}}\delta _{b\bar{c}},\]
\[
T_{7}=if^{n_{1}n_{2}g_{2}}t^{n_{1}}_{c\bar{c}}t_{bl_{1}}^{g_{1}}t_{l_{1}\bar{b}}^{n_{2}}\delta _{b\bar{c}},\]
\[
T_{8}=if^{n_{1}n_{2}g_{2}}t^{n_{1}}_{c\bar{c}}t_{bl_{1}}^{n_{2}}t_{l_{1}\bar{b}}^{g_{1}}\delta _{b\bar{c}},\]
\[
T_{9}=if^{n_{1}n_{2}g_{2}}t_{b\bar{b}}^{n_{1}}t_{cl_{1}}^{g_{1}}t_{l_{1}\bar{c}}^{n_{2}}\delta _{b\bar{c}},\]
\[
T_{10}=if^{n_{1}n_{2}g_{2}}t_{b\bar{b}}^{n_{1}}t_{cl_{1}}^{n_{2}}t_{l_{1}\bar{c}}^{g_{1}}\delta _{b\bar{c}},\]
\[
T_{11}=if^{n_{1}n_{2}g_{1}}t_{b\bar{b}}^{n_{1}}t_{cl_{1}}^{n_{2}}t_{l_{1}\bar{c}}^{g_{2}}\delta _{b\bar{c}},\]
\[
T_{12}=if^{n_{1}n_{2}g_{1}}t_{b\bar{b}}^{n_{1}}t_{cl_{1}}^{g_{2}}t_{l_{1}\bar{c}}^{n_{2}}\delta _{b\bar{c}},\]
 \[
T_{13}=if^{n_{1}n_{2}g_{1}}t^{n_{1}}_{c\bar{c}}t_{bl_{1}}^{n_{2}}t_{l_{1}\bar{b}}^{g_{2}}\delta _{b\bar{c}},\]
\[
T_{14}=if^{n_{1}n_{2}g_{1}}t^{n_{1}}_{c\bar{c}}t_{bl_{1}}^{g_{2}}t_{l_{1}\bar{b}}^{n_{2}}\delta _{b\bar{c}},\]
\[
T_{15}=t_{bl_{1}}^{g_{1}}t_{l_{1}\bar{b}}^{n_{1}}t_{cl_{2}}^{n_{1}}t_{l_{2}\bar{c}}^{g_{2}}\delta _{b\bar{c}},\]
\[
T_{16}=t_{bl_{1}}^{n_{1}}t_{l_{1}\bar{b}}^{g_{1}}t_{cl_{2}}^{n_{1}}t_{l_{2}\bar{c}}^{g_{2}}\delta _{b\bar{c}},\]
\[
T_{17}=t_{bl_{1}}^{g_{1}}t_{l_{1}\bar{b}}^{n_{1}}t_{cl_{2}}^{g_{2}}t_{l_{2}\bar{c}}^{n_{1}}\delta _{b\bar{c}},\]
\[
T_{18}=t_{bl_{1}}^{g_{1}}t_{l_{1}\bar{b}}^{n_{1}}t_{cl_{2}}^{g_{2}}t_{l_{2}\bar{c}}^{n_{1}}\delta _{b\bar{c}},\]
\[
T_{19}=t_{bl_{1}}^{n_{1}}t_{l_{1}\bar{b}}^{g_{2}}t_{cl_{2}}^{g_{1}}t_{l_{2}\bar{c}}^{n_{1}}\delta _{b\bar{c}},\]
\[
T_{20}=t_{bl_{1}}^{n_{1}}t_{l_{1}\bar{b}}^{g_{2}}t_{cl_{2}}^{n_{1}}t_{l_{2}\bar{c}}^{g_{1}}\delta _{b\bar{c}},\]
\[
T_{21}=t_{bl_{1}}^{n_{1}}t_{l_{1}\bar{b}}^{g_{2}}t_{cl_{2}}^{n_{1}}t_{l_{2}\bar{c}}^{g_{1}}\delta _{b\bar{c}},\]
\[
T_{22}=t_{bl_{1}}^{g_{2}}t_{l_{1}\bar{b}}^{n_{1}}t_{cl_{2}}^{n_{1}}t_{l_{2}\bar{c}}^{g_{1}}\delta _{b\bar{c}},\]
\[
T_{23}=t_{bl_{1}}^{n_{1}}t_{l_{1}l_{2}}^{g_{1}}t_{l_{2}\bar{b}}^{g_{2}}t_{c\bar{c}}^{n_{1}}\delta _{b\bar{c}},\]
\[
T_{24}=t_{bl_{1}}^{g_{1}}t_{l_{1}l_{2}}^{n_{1}}t_{l_{2}\bar{b}}^{g_{2}}t_{c\bar{c}}^{n_{1}}\delta _{b\bar{c}},\]
\[
T_{25}=t_{bl_{1}}^{g_{1}}t_{l_{1}l_{2}}^{g_{2}}t_{l_{2}\bar{b}}^{n_{1}}t_{c\bar{c}}^{n_{1}}\delta _{b\bar{c}},\]
\[
T_{26}=t_{bl_{1}}^{n_{1}}t_{l_{1}l_{2}}^{g_{2}}t_{l_{2}\bar{b}}^{g_{1}}t_{c\bar{c}}^{n_{1}}\delta _{b\bar{c}},\]
\[
T_{27}=t_{bl_{1}}^{g_{2}}t_{l_{1}l_{2}}^{n_{1}}t_{l_{2}\bar{b}}^{g_{1}}t_{c\bar{c}}^{n_{1}}\delta _{b\bar{c}},\]
\[
T_{28}=t_{bl_{1}}^{g_{2}}t_{l_{1}l_{2}}^{g_{1}}t_{l_{2}\bar{b}}^{n_{1}}t_{c\bar{c}}^{n_{1}}\delta _{b\bar{c}},\]
\[
T_{29}=t_{cl_{1}}^{n_{1}}t_{l_{1}l_{2}}^{g_{1}}t_{l_{2}\bar{c}}^{g_{2}}t_{b\bar{b}}^{n_{1}}\delta _{b\bar{c}},\]
\[
T_{30}=t_{cl_{1}}^{g_{1}}t_{l_{1}l_{2}}^{n_{1}}t_{l_{2}\bar{c}}^{g_{2}}t_{b\bar{b}}^{n_{1}}\delta _{b\bar{c}},\]
 \[
T_{31}=t_{cl_{1}}^{g_{1}}t_{l_{1}l_{2}}^{g_{2}}t_{l_{2}\bar{c}}^{n_{1}}t_{b\bar{b}}^{n_{1}}\delta _{b\bar{c}},\]
\[
T_{32}=t_{cl_{1}}^{n_{1}}t_{l_{1}l_{2}}^{g_{2}}t_{l_{2}\bar{c}}^{g_{1}}t_{b\bar{b}}^{n_{1}}\delta _{b\bar{c}},\]
 \[
T_{33}=t_{cl_{1}}^{g_{2}}t_{l_{1}l_{2}}^{n_{1}}t_{l_{2}\bar{c}}^{g_{1}}t_{b\bar{b}}^{n_{1}}\delta _{b\bar{c}},\]
\[
T_{34}=t_{cl_{1}}^{g_{2}}t_{l_{1}l_{2}}^{g_{1}}t_{l_{2}\bar{c}}^{n_{1}}t_{b\bar{b}}^{n_{1}}\delta _{b\bar{c}},\]
\[
T_{35}=if^{n_{1}g_{1}g_{2}}t_{bl_{1}}^{n_{2}}t_{l_{1}\bar{b}}^{n_{1}}t_{c\bar{c}}^{n_{2}}\delta _{b\bar{c}},\]
\[
T_{36}=if^{n_{1}g_{1}g_{2}}t_{bl_{1}}^{n_{1}}t_{l_{1}\bar{b}}^{n_{2}}t_{c\bar{c}}^{n_{2}}\delta _{b\bar{c}},\]
\[
T_{37}=if^{n_{1}g_{1}g_{2}}t_{cl_{1}}^{n_{2}}t_{l_{1}\bar{c}}^{n_{1}}t_{b\bar{b}}^{n_{2}}\delta _{b\bar{c}},\]
\[
T_{38}=if^{n_{1}g_{1}g_{2}}t_{cl_{1}}^{n_{1}}t_{l_{1}\bar{c}}^{n_{2}}t_{b\bar{b}}^{n_{2}}\delta _{b\bar{c}},\]
where  upper indexes  \( g_{1} \), \( g_{2} \) are color states of the 
initial gluons, lower indexes \( b \), \( \bar{b} \),
\( c \), \( \bar{c} \) are color states of \( b \)-, \( \bar{b} \)-,
\( c \)-, \( \bar{c} \)-quarks correspondingly 
and  \( \delta _{b\bar{c}} \) is a color wave function of the
 \( B_{c} \)-meson  ( the normalization coefficient 
 \( 1/\sqrt{3} \) is not written for the sake of  simplicity).

Let us study, for example, the color part of diagram 1
(see  Fig.~\ref{diagr}. In this diagram the initial guons
 exchange a gluon in t-channel and split into
the quark-antiquark pairs):

\[
T_{1}=f^{n_{1}g_{2}n_{2}}f^{n_{3}n_{2}g_{1}}t_{c\bar{c}}^{n_{1}}t_{b\bar{b}}^{n_{3}}\delta _{b\bar{c}}=f^{n_{1}g_{2}n_{2}}f^{n_{3}n_{2}g_{1}}(t^{n_{1}}t^{n_{3}})_{\bar{b}c}.\]

Using the identity \( t^{a}t^{b}-t^{b}t^{a}=if^{abc}t^{c}, \) one can found:

\[
f^{n_{1}g_{2}n_{2}}f^{n_{3}n_{2}g_{1}}t^{n_{1}}t^{n_{3}}=-(t^{g_{2}}t^{n_{2}}-t^{n_{2}}t^{g_{2}})(t^{n_{2}}t^{g_{1}}-t^{g_{1}}t^{n_{1}})=\]

\[
=-t^{g_{2}}t^{n_{2}}t^{n_{2}}t^{g_{1}}+t^{n_{2}}t^{g_{2}}t^{n_{2}}t^{g_{1}}+t^{g_{2}}t^{n_{2}}t^{g_{1}}t^{n_{1}}-t^{n_{2}}t^{g_{2}}t^{g_{1}}t^{n_{1}}=\]

\[
-\frac{4}{3}t^{g_{2}}t^{g_{1}}-\frac{1}{6}t^{g_{2}}t^{g_{1}}-\frac{1}{6}t^{g_{2}}t^{g_{1}}-(\frac{1}{4}\delta ^{g_{1}g_{2}}-\frac{1}{6}t^{g_{2}}t^{g_{1}})=\]

\[
=-\frac{1}{4}\delta ^{g_{1}g_{2}}-\frac{3}{2}t^{g_{2}}t^{g_{1}}.\]

Thus:

\[
T_{1}=-\frac{3}{2}t_{ck}^{g_{2}}t^{g_{1}}_{k\bar{b}}-\frac{1}{4}\delta ^{g_{1}g_{2}}\delta _{c\bar{b}}.\]

\begin{figure}
{\centering \resizebox*{0.8\textwidth}{!}{\includegraphics{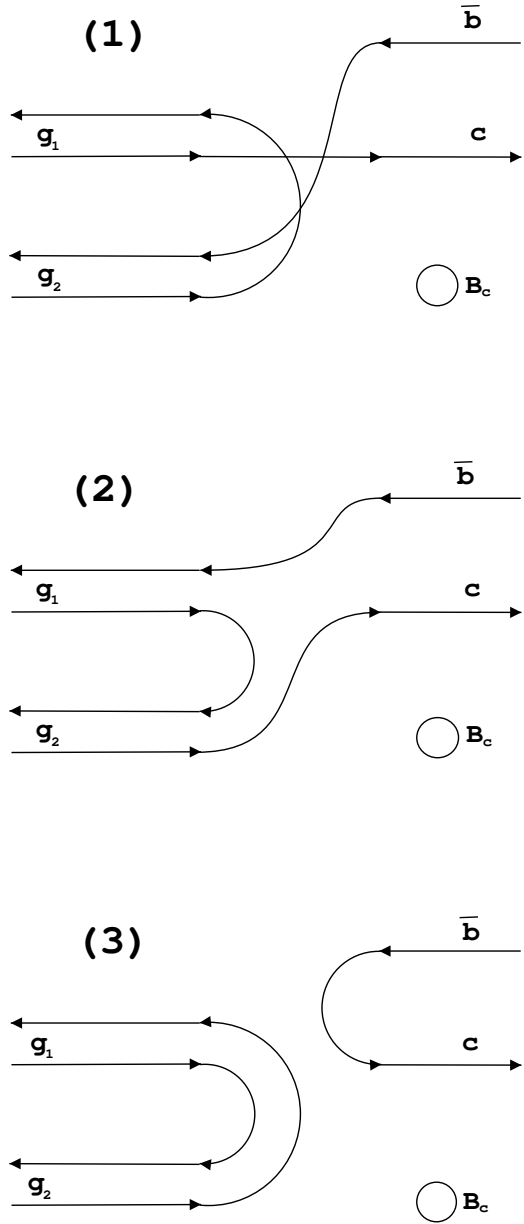}} \par}

\setcaptionmargin{5mm}
\captionstyle{normal}
\caption{
Color flows for the process 
\protect\( gg\rightarrow B_{c}+c+\bar{b}\protect \). 
Color strings stretch as follows for the color flow (1): 
\protect\([c\protect\)-quark 
\protect\(\longrightarrow\protect\)
the remnant of hadron, which contained the gluon \protect\(g_1\protect\) 
\protect\(\longrightarrow\protect\)
the remnant of hadron, which contained the gluon  \protect\(g_2\protect\) 
\protect\(\longrightarrow\protect\)
\protect\(\bar{b}\protect\)-quark\protect\(]\protect\); 
for the color flow (2): 
\protect\([c\protect\)-quark 
\protect\(\longrightarrow\protect\) 
the remnant of hadron, which contained the gluon  \protect\(g_2\protect\) 
\protect\(\longrightarrow\protect\)
the remnant of hadron, which contained the gluon \protect\(g_1\protect\) 
\protect\(\longrightarrow\protect\)
\protect\(\bar{b}\protect\)-quark\protect\(]\protect\);  
for the color flow (3):  
\protect\([c\protect\)-quark 
\protect\(\longrightarrow\protect\)
\protect\(\bar{b}\protect\)-quark\protect\(]\protect\), 
\protect\([\protect\)the remnant of hadron, 
which contained the gluon \protect\(g_1\protect\) 
\protect\(\rightleftarrows\protect\)
the remnant of hadron, 
which contained the  gluon  \protect\(g_2]\protect\).\hfill
\label{colors}}
\end{figure}

The first term  (see the scheme (2) in Fig.~\ref{colors}) corresponds to
the case, where a color of the second gluon  (\( g_{2} \) 
 flows to \( c \)-quark,
an anticolor of the first gluon  (\( g_{1} \)) flows  to \( \bar{b} \)-quark,
and a color of the first gluon annihilates with an anticolor of 
the second gluon (sum over \( k \)).
The second term corresponds to
the case, where  colors and anticolors of the initial gluons annihilate,
and a color and an anticolor of 
\( c \)-quark  and \( \bar{b} \)-quark are produced from a vacuum
 (see the scheme (3) in Fig.~\ref{colors}). 
In addition to the terms described above, the term  
\( t_{ck}^{g_{1}}t_{k\bar{b}}^{g_{2}} \) 
contributes to another diagrams. The latter one corresponds to
the case, 
where a color of the first gluon  (\( g_{1} \)  flows to \( c \)-quark,
an anticolor of the second gluon  (\( g_{2} \)) flows  
to \( \bar{b} \)-quark,
and an anticolor of the first gluon annihilates with an color of 
the second gluon (see the scheme (1) in Fig.~\ref{colors}). 
There are no  color flows for discussed  process but such as 
three ones described above. 
This color flow separation  has been done by following the 
recipe given in papers  \cite{c_string}. 

Nevertheless, it would be better to base on  more 
fundamental QCD principles to describe color flows.
Let us consider the term  \( \delta ^{g_{1}g_{2}}\delta _{c\bar{b}} \),
which corresponds to the production of 
\( c \)-quark and \( \bar{b} \)-quark in  a color singlet.
Naturally, a color sting stretch  between these two quarks.
It is worth to note that the term \( t_{ck}^{g_{2}}t^{g_{1}}_{k\bar{b}} \) 
contains the singlet part too:
\begin{equation}
\label{t_f_d}
t^{a}t^{b}=\frac{1}{6}\delta ^{ab}+\frac{1}{2}
(d^{abc}+if^{abc})t^{c}.
\end{equation}

Therefore "one part" of the singlet is hadronized in one manner,
and "other part" of the singlet is hadronized in other manner.
We think that it would be better to treat 
the total color singlet contribution as the separate color flow.
Two other color flows would be composed of two color octet states 
\( d \) and \( f \).

That is why we redefine the color flows as follows:
\begin{enumerate}
\item A color of the first gluon flows to \( c \)-quark, an anticolor
of the second gluon flows to \( \bar{b} \)-quark, 
an anticolor of the first gluon and a color of the second one
annihilate: \[
\frac{1}{2}(d^{g_{1}g_{2}k}+if^{g_{1}g_{2}k})t^{k}_{c\bar{b}}.\]

\item A color of the second gluon flows to \( c \)-quark, an anticolor
of the first gluon flows to \( \bar{b} \)-quark, 
a color of the first gluon and an anticolor of the second one
annihilate: \[
\frac{1}{2}(d^{g_{1}g_{2}k}-if^{g_{1}g_{2}k})t^{k}_{c\bar{b}}.\]

\item A color and an anticolor of the initial gluons annihilate,
a color of \( c \)-quark and an anticolor of 
 \( \bar{b} \)-quark are produced from vacuum:\[
\delta ^{g_{1}g_{2}}\delta _{c\bar{b}}.\]

\end{enumerate}
In our point of view these definitions of colors flows are more
physically justified, because the singlet state contribution is separated from 
the octet contributions and 
there is no an interference term between the singlet
color flow and other flows.

The definitions described above slightly differ from 
the traditional ones: \( t_{ck}^{g_{1}}t^{g_{2}}_{k\bar{b}} \),
\( t_{ck}^{g_{2}}t^{g_{1}}_{k\bar{b}} \) 
and \( \delta ^{g_{1}g_{2}}\delta _{c\bar{b}} \).
However, an accuracy of the color flow separation is about  
\( 1/N_{c} \),
where \( N_{c} \) is a color number. For an arbitrary \( N_{c} \) the 
equation (\ref{t_f_d}) looks like follows:
\begin{equation}
\label{t_f_d_N}
t^{a}t^{b}=
\frac{1}{2N_{c}}\delta ^{ab}+\frac{1}{2}(d^{abc}+if^{abc})t^{c},
\end{equation}
and one can conclude that at \( 1/N_{c}\rightarrow \infty  \), the both 
definition sets lead to the same results. 

Thus the color part of matrix element  \( n \) can be given by
the following expression:

\begin{equation}
\label{my_cf}
T_{n}=\frac{1}{2}(d^{g_{1}g_{2}k}+if^{g_{1}g_{2}k})t^{k}_{c\bar{b}}\cdot A_{n}+\frac{1}{2}(d^{g_{1}g_{2}k}-if^{g_{1}g_{2}k})t^{k}_{c\bar{b}}\cdot B_{n}+\delta ^{g_{1}g_{2}}\delta _{c\bar{b}}\cdot C_{n},
\end{equation}

From (\ref{my_cf}) one can easy obtain 
the color matrix averaged over the initial color states 
and summed over the final ones:
\begin{equation}
\label{c_matr}
M_{mn}=\frac{1}{64}\Bigl ((D+F)\cdot A_{m}A_{n}+(D+F)\cdot B_{m}B_{n}+(D-F)\cdot (A_{m}B_{n}+B_{m}A_{n})+S\cdot C_{m}C_{n}\Bigr ),
\end{equation}
where \( D=5/3 \), \( F=3 \), and \( S=24 \). 
Vectors  \( A_{n} \), \( B_{n} \) and  \( C_{n} \) 
are performed at the table \ref{c_tab}.

In our approach the color matrix corresponded with an interference between 
the color flows \( (d^{g_{1}g_{2}k}+if^{g_{1}g_{2}k})t^{k}_{c\bar{b}} \)
and \( (d^{g_{1}g_{2}k}-if^{g_{1}g_{2}k})t^{k}_{c\bar{b}} \) 
have a simple form:
\begin{equation}
\label{c_matr_int}
M^{\rm int}_{mn}=\frac{1}{64}\Bigl ((D-F)\cdot (A_{m}B_{n}+B_{m}A_{n})\Bigr )
\end{equation}

It is worth to note, that the formula (\ref{my_cf}) can be easily rewritten
in the traditional color flow definitions:

\begin{equation}
\label{old_cf}
T_{n}=t_{ck}^{g_{1}}t^{g_{2}}_{k\bar{b}}\cdot A_{n}+t_{ck}^{g_{2}}t^{g_{1}}_{k\bar{b}}\cdot B_{n}+\delta ^{g_{1}g_{2}}\delta _{c\bar{b}}\cdot (C_{n}-\frac{A_{n}+B_{n}}{6}).
\end{equation}

One can see that the expressions for  the color flow (1) 
given by  (\ref{my_cf}) and  (\ref{old_cf}) differ from each other only
in a common coefficient.  It is easily to show that for the color flow (1)
the ratio for the matrix element from (\ref{my_cf})  to one from 
(\ref{old_cf}) is \( 7:8 \), as well as for (2).

\section{CALCULATION RESULTS}

\begin{figure}
{\centering \resizebox*{0.8\textwidth}{!}{\includegraphics{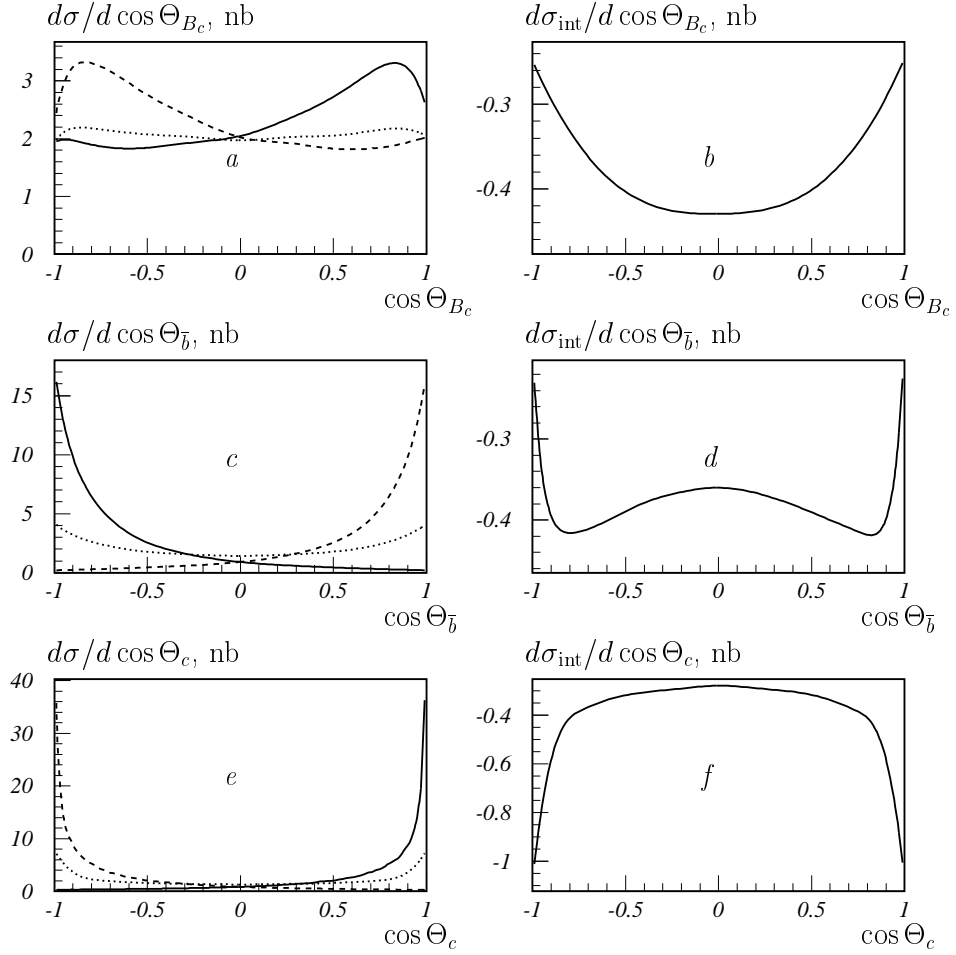}} \vspace*{-4cm}\par}

\setcaptionmargin{5mm}
\onelinecaptionsfalse
\captionstyle{normal}
\caption{The cross section distributions for 
the pseudoscalar \protect\( B_c \protect\)-meson
production at 25 GeV (\protect\( gg\rightarrow B_{c}+\bar{b}+c\protect \))
for the different color flows and the interference term:
over the exit angle of \protect\( B_{c}\protect \)-meson ( a) and b) );
over the exit angle of \protect\( \bar{b}\protect \)-quark ( c) and d) ); 
over the exit angle of \protect\( c\protect \)-quark ( e) and f) ). 
A solid curve in plots a), c) and e) denotes the color flow (1); 
a dashed curve denotes the color flow (2); a dotted one denotes 
the color flow (3). The interference contribution is shown 
in plots b), d) and f). 
\hfill \label{pseudoscalar}}
\end{figure}

\begin{figure}
{\centering \resizebox*{0.8\textwidth}{!}{\includegraphics{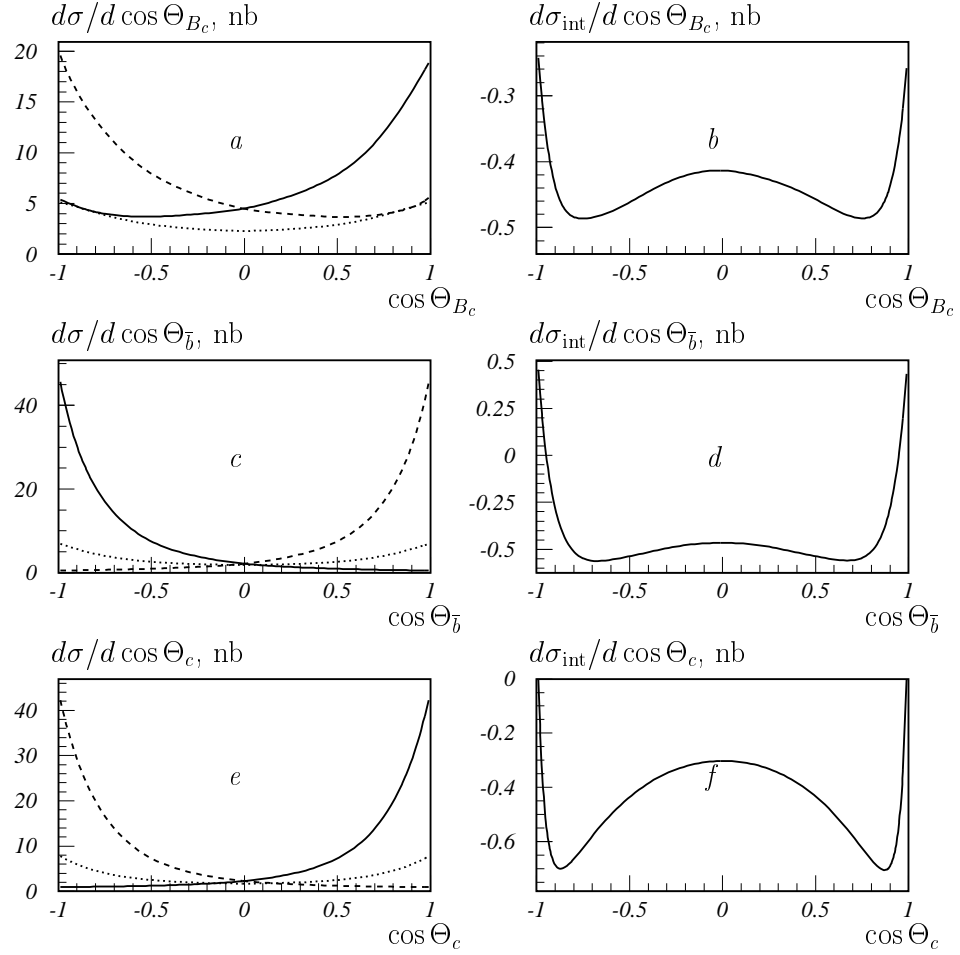}} 
\vspace*{-4cm}\par}

\setcaptionmargin{5mm}
\onelinecaptionsfalse
\captionstyle{normal}
\caption{The cross section distributions for the
vector \protect\( B_c \protect\)-meson
production at 25 GeV for the different color flows
  and the interference term. 
The  designations are the same as in Fig.~\ref{pseudoscalar}.
\hfill\label{vector}\hfill}
\end{figure}

The cross section distributions over exit angles of final particles 
(\( B_{c} \), \( \bar{b} \) and \( c \))
have been shown in Fig.~\ref{pseudoscalar} and \ref{vector}
for the different color flows and for the interference term at the gluon 
interaction energy 25 GeV. This energy value has been chosen because 
it is common  value  for  \( B_{c} \)-meson production at LHC. It is worth to
mention that intuitive ideas about color flows correspond to the calculation
results. Indeed, it is clear from Fig.~\ref{pseudoscalar}c and \ref{vector}c
that \( \bar{b} \)-quark moves mainly  in the direction 
of  gluon, which transfer an anticolor to \( \bar{b} \)-quark,
Also one can see from Fig.~\ref{pseudoscalar}f and \ref{vector}f, 
that  \( c \)-quark   moves mainly  in the direction 
of  gluon, which transfer a color to \( c \)-quark.
It is not unexpected, that the  color flow (3),
which   corresponds to a singlet state, 
is symmetrically distributed over the angles.

The interference term between flows 
(1) and (2) is small (see Fig.~\ref{pseudoscalar}b,
d, f É \ref{vector}b, d, f). For the pseudoscalar  \( B_{c} \)-meson
production at 25 GeV the interference contribution is negative 
for all values of the exit angles.  For the vector meson
production the interference contribution is negative too
with the exception of peripheral regions of the angular distributions.
The total interference contribution into the vector meson
production is negative, as well as into the pseudoscalar meson production. 
Our calculations show that at low energies the  interference contribution  
becomes positive for the vector meson production  and remains negative for
the pseudoscalar meson production.  An absolute value of the
interference contribution is small at all interaction energies. 

\begin{figure}
{\centering \resizebox*{0.8\textwidth}{!}{\includegraphics{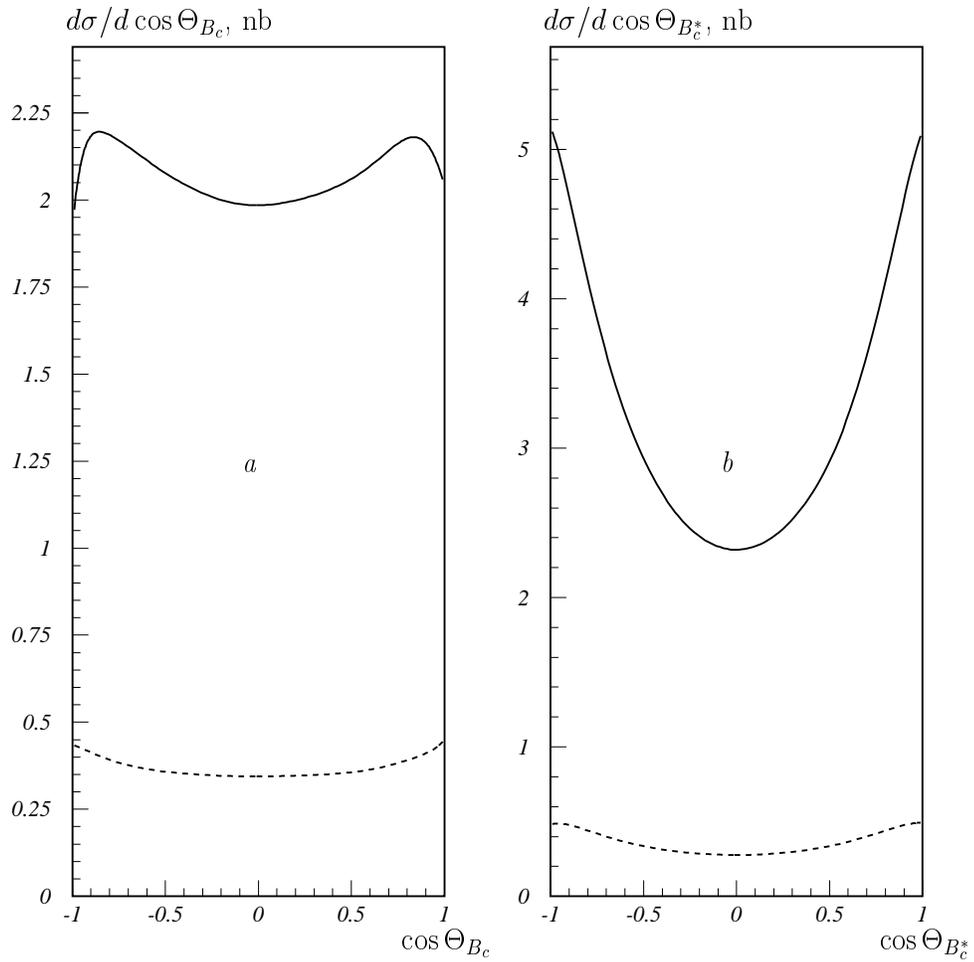}} 
\vspace*{-4cm}
\par}

\setcaptionmargin{5mm}
\onelinecaptionsfalse
\captionstyle{normal}
\caption{The cross section distributions over
the exit angle of pseudoscalar (a) and vector (b) 
\protect\( B_{c}\protect \)-meson for the color flow (3):
in our approach (solid curve)
 and in the traditional one (dashed curve). \hfill \label{compare} }
\end{figure}

As it was mentioned above, 
the  contributions of the color flows (1) and (2) in our separation scheme
differ from the traditionally determined  contributions only by
the common coefficient. It is not so for the color flow (3). 
The contribution of (3) in our approach one in traditional 
approach differs from each other in shape too. The distributions over 
exit angle of  \( B_{c} \)-meson in the frame work of traditional 
and our approaches have been presented in  Fig. \ref{compare}. 
One can see that in our approach the contribution of (3) is
larger than this contribution in the traditional scheme.
The point is that in our approach 
there is no interference between the color flows (1) and (3), as well as
between  (2) and (3). Therefore the contributions, which correspond
in traditional scheme to the  interference between (1) and (3), as well as
between  (2) and (3), transfer to color flow (3).

\section{CONCLUSIONS}

The proposed calculation method 
 minimize interference terms between color flows
for the process \( gg\rightarrow B_{c}+\bar{b}+c\rightarrow B_{c}+X \).
That is why interference terms can be neglected in our approach.
Furthermore the method under discussion allows us to ascertain more clearly
how color states of final partons correspond with color flows.
The redistribution of interference terms leads to the amplification of the 
color flow (3). The contribution of this color flow into the central kinematic
region becomes  comparable with the contributions of color flows (1) and (2). 
Thus haronization features of the outgoing partons \( \bar{b} \) and \( c \) 
could be changed. 

\begin{acknowledgments}
I thank  A.~K.~Likhoded for the fruitful discussion of 
this paper. Also I am grateful to Y.~P.~Gouz and
S.~G.~Shulga for the collaboration  
in  \( B_{c} \) production simulation at LHC.
This research is partially supported  by 
Grants of RF Education Ministry   å02-3.1-96,
CRDF  M0-011-0 and RFBR  04-02-17530, and is realized 
in the frame work of scientific school  SC~1303.2003.2.

\end{acknowledgments}

\newpage

\begin{table}

\setcaptionmargin{5mm}
\onelinecaptionsfalse
\captionstyle{normal}
\caption{The vectors 
\protect\( A_{n}\protect \) , \protect\( B_{n}\protect \),
\protect\( C_{n}\protect \)  corresponded
with the color flows  (1), (2), (3) in  Fig.~\ref{colors}.\hfill\label{c_tab}}

\vspace*{0.5cm}
\centering
\begin{tabular}{|c|c|c|c|}
\hline 
\( n \)&
\( A_{n} \)&
\( B_{n} \)&
\( C_{n} \)\\
\hline
\hline 
1&
0&
\( -3/2 \)&
\( -1/2 \)\\
\hline 
2&
\( -3/2 \) &
0&
\( -1/2 \)\\
\hline 
3&
\( 3/2 \)&
-\( 3/2 \)&
0\\
\hline 
4&
0&
\( 3/2 \)&
\( 1/2 \)\\
\hline 
5&
\( -3/2 \)&
0&
\( -1/2 \)\\
\hline 
6&
\( 3/2 \) &
\( -3/2 \)&
0\\
\hline 
7&
0&
0&
\( 1/4 \)\\
\hline 
8&
0&
-\( 3/2 \)&
\( -1/4 \)\\
\hline 
9&
\( 3/2 \)&
0&
\( 1/4 \)\\
\hline 
10&
0&
0&
\( -1/4 \)\\
\hline 
11&
0&
0&
\( -1/4 \)\\
\hline 
12&
0&
\( 3/2 \)&
\( 1/4 \)\\
\hline 
13&
\( -3/2 \)&
0&
\( -1/4 \)\\
\hline 
14&
0&
0&
\( 1/4 \)\\
\hline 
15&
0&
\( -1/6 \)&
\( 2/9 \)\\
\hline 
16&
0&
\( -1/6 \)&
\( -1/36 \)\\
\hline 
17&
0&
\( -1/6 \)&
\( -1/36 \)\\
\hline 
18&
0&
\( 4/3 \)&
\( 2/9 \)\\
\hline
19&
\( 4/3 \)&
0&
\( 2/9 \)\\
\hline 
\end{tabular}
\hspace{0.5cm}
\begin{tabular}{|c|c|c|c|}
\hline 
\( n \)&
\( A_{n} \)&
\( B_{n} \)&
\( C_{n} \)\\
\hline
\hline 
20&
\( -1/6 \)&
0&
\( -1/36 \)\\
\hline 
21&
\( -1/6 \)&
0&
\( -1/36 \)\\
\hline 
22&
\( -1/6 \)&
0&
\( 2/9 \)\\
\hline 
23&
\( 4/3 \)&
0&
\( 2/9 \)\\
\hline 
24&
\( -1/6 \)&
0&
\( -1/36 \)\\
\hline 
25&
\( -1/6 \)&
0&
\( 2/9 \)\\
\hline 
26&
0&
\( 4/3 \)&
\( 2/9 \)\\
\hline 
27&
0&
\( -1/6 \)&
\( -1/36 \)\\
\hline 
28&
0&
\( -1/6 \)&
\( 2/9 \)\\
\hline 
29&
\( -1/6 \)&
0&
\( 2/9 \)\\
\hline 
30&
\( -1/6 \)&
0&
\( -1/36 \)\\
\hline 
31&
\( 4/3 \)&
0&
\( 2/9 \)\\
\hline 
32&
0&
\( -1/6 \)&
\( 2/9 \)\\
\hline 
33&
0&
\( -1/6 \)&
\( -1/36 \)\\
\hline 
34&
0&
\( 4/3 \)&
\( 2/9 \)\\
\hline 
35&
\( 4/3 \)&
\( -4/3 \)&
0\\
\hline 
36&
\( -1/6 \)&
\( 1/6 \)&
0\\
\hline 
37&
\( -1/6 \)&
\( 1/6 \)&
0\\
\hline 
38&
\( 4/3 \)&
\( -4/3 \)&
0\\
\hline
\end{tabular}
\end{table}

\end{document}